\documentclass[a4paper,fleqn,usenatbib]{mnras}
 
\usepackage{newtxtext,newtxmath}

\usepackage[T1]{fontenc}
\usepackage{ae,aecompl}

\usepackage{graphicx}	
\usepackage{amssymb}	

\title[$M/R$ estimates for neutron stars in LMXBs]{$M/R$ estimates for two neutron stars in LMXBs with possible r-mode frequencies detected}

\author[C. Chirenti and M. Jasiulek]{
Cecilia Chirenti,$^{1}$\thanks{E-mail: cecilia.chirenti@ufabc.edu.br}
and Michael Jasiulek$^{1}$
\\
$^{1}$Center for Mathematics, Computation and Cognition, UFABC, Av. dos Estados, 5001, Santo Andr\'e - SP, 09210-580, Brazil
}

\date{Accepted XXX. Received YYY; in original form ZZZ}

\pubyear{2017}

\begin{document}
\label{firstpage}
\pagerange{\pageref{firstpage}--\pageref{lastpage}}
\maketitle

\begin{abstract}
The puzzling existence of a number of neutron stars that appear to be in the r-mode instability window supports further investigations of the r-mode frequencies, damping times and saturation amplitudes, especially now in light of the much anticipated and exciting results from LIGO and NICER. It has been suggested by Mahmoodifar and Strohmayer that coherent frequencies found in the RXTE data of the accreting millisecond pulsar XTE J1751-305 and during the 2001 superburst of 4U 1636-536 (one of the accreting sources to be observed by NICER) could, in fact, be r-modes. Based on these observations, we expand here the results of a previous work on relativistic, rotational and differential rotation corrections to the r-mode frequency in order to provide more accurate estimates of the $M-R$ relation for these neutron stars. Finally, we compare our results with predictions from a few realistic tabulated equations of state, providing further constraints that favor more compact models. We find that, if the observed frequencies indeed correspond to r-modes, then the masses of XTE J1751-305 and 4U 1636-536 should lie in the approximate ranges  $1.48 - 1.56 M_{\odot}$  and $1.60 - 1.68 M_{\odot}$, respectively.
\end{abstract}




\section{Introduction}
\label{sec:intro}

It is an exciting time to work with neutron stars and gravitational waves. After the confirmed detection of 4 binary black hole mergers \citep{PhysRevLett.116.061102,PhysRevLett.116.241103,PhysRevLett.118.221101,PhysRevLett.119.141101}, the first detection of gravitational waves from a binary neutron star merger by LIGO and Virgo has initiated the age of multimessenger astronomy with gravitational waves and it has already provided some preliminary constraints on the neutron star equation of state \citep{PhysRevLett.119.161101}. Moreover, the launch of the NICER mission last July \citep{2017HEAD...2771250} increases the expectations for the determination of masses and radii of the primary target neutron stars with both high precision and accuracy.

Neutron stars have long been proposed as sources of gravitational waves, but not only in binaries. Oscillation modes of rotating neutron stars are subject to the CFS instability \citep{1970ApJ...161..561C,1978ApJ...222..281F} and r-modes, in particular, are unstable even for slowly rotating stars. The r-modes are coupled to the emission of gravitational waves, and their amplitude will increase until it reaches a certain saturation amplitude, which is still the topic of much debate. The comparison between the r-mode growth timescale and the stellar fluid viscosity damping timescale shapes the instability window in the period-temperature diagram.

According to the usual picture, the r-mode instability will be activated if a neutron star happens to enter the instability window, and then the star will quickly radiate gravitational waves that will take away its extra angular momentum until the star is once again out of the instability window. This mechanism has also been proposed as a possible explanation for the observed periods of pulsars. However, it is a puzzling observational fact that there are a few neutron stars that seem to be inside the window \citep{2017ApJ...840...94M}. More precise calculations of the r-modes and their instability window are needed in order to help explain these observations. 

Another interesting piece of the puzzle comes from the observation by Mahmoodifar and Strohmayer of coherent oscillations in the spectra of XTE J1751-305 \citep{2014ApJ...784...72S} and 4U 1636-536 \citep{2014ApJ...793L..38S}. Their frequencies are tantalizingly close to the Newtonian estimate for the r-mode frequency of these stars. 

\cite{2014ApJ...784...72S} suggested that possible identifications of the observed frequency include rotationally modified g-modes or alternatively an inertial mode or core r-mode modified by the presence of a solid crust. However, the r-mode amplitude derived from the observed modulation amplitude would lead to a spin-down rate too large to be consistent with the observations. The spin evolution of XTE J1751-305 was rediscussed by \citet{2014MNRAS.442.1786A}, who concluded from their analysis of different r-mode scenarios (stable/unstable and saturated/unsaturated) that the presence of an r-mode with the large amplitude suggested by \citet{2014ApJ...784...72S} cannot be reconciled with the observed spin-up of the star during the outburst and must therefore be considered unlikely.
However, it has also been suggested by \cite{2014MNRAS.442.3037L} that the frequency could be a \emph{surface} g-mode or r-mode, and alternatives like unstable toroidal crust modes or the amplification at the surface of a core r-mode with a lower amplitude were also discussed.

In a previous work \citep{2017PhRvD..95f4060J} we have carefully calculated the corrections to the r-mode frequencies taking into account relativist and rotational effects in fast rotating neutron stars, also including a possibly strong differential rotation. These results were obtained in the relativistic Cowling approximation (that neglects all metric perturbations), by numerically solving the linear perturbation equations in the fixed stellar background and extending to fast rotation our previous results calculated in the slow rotation approximation \citep{2013PhRvD..87d4043C,2013PhRvD..88j4018C}.

In the present paper we improve on our previous results by proposing a method to estimate the error due to the Cowling approximation, and extending our calculations for stars with different radii. We use our improved results to estimate allowed mass ranges for XTE J1751-305 and 4U 1636-536, given a representative range for the neutron stars' radius \citep{2016EPJA...52...63M}, and allowing the stars to have a small differential rotation (compatible with what is expected from glitch models \citep{2014ApJ...789..141L}). Our analysis is an improvement of the work presented in \citet{2014MNRAS.442.1786A}, which used as a first approximation the fully relativistic calculation of the r-mode frequency for slowly rotating uniform density stars from \citet{2003PhRvD..68l4010L} with the second order rotational corrections calculated for fast rotating Newtonian polytropes given in \citet{1999PhRvD..60f4006L}.

This paper is organized as follows. In Section \ref{sec:M-R} we briefly discuss some of the current expectations for the determination of neutron stars masses and radii, and how the r-mode observations can also be used for this purpose.  The most relevant corrections to the r-mode frequency are reviewed in Section \ref{sec:rmode}. We present our numerical results for the mass-radius constraints in Section \ref{sec:results} and our final conclusions in Section \ref{sec:conclusion}.

\section{Determination of neutron star masses and radii}
\label{sec:M-R}

\subsection{Current data and expectations}
\label{sec:M-Rdata}

Determinations of neutron star masses in binary systems have provided the most precise and reliable mass estimates so far \citep{2016EPJA...52...63M}. Current observations indicate a maximum mass of at least 2 $M_{\odot}$, with mass distributions for different populations of neutron stars showing a possible bimodal distribution with peaks at $\approx 1.3 M_{\odot}$ and $\approx 1.5 M_{\odot}$ \citep{2016ARA&A..54..401O}. Different models indicate that the neutron star radius could be as low as $\approx$ 10 km or as high as $\approx$ 15 km, but several models predict a radius from $\approx$ 11 to $\approx$ 13 km (see for instance fig. 2 of \citet{2012PhRvC..85c2801G}, or figs. 10 and 9 of \citet{2014EPJA...50...40L,2014ApJ...784..123L}, respectively). We will take this as our representative range for the neutron star's radius. Precise determinations of the mass $M$ and radius $R$ for several neutron stars would be enormously helpful to rule out models and equations of state, but systematic errors could render results unreliable (even if they are formally statistically precise) \citep{2016EPJA...52...63M}.

The NICER mission is expected to provide both precise and accurate measurements of $(M,R)$ for 4 non-accreting main target stars. Right now, NICER is already installed on the international space station and taking data \citep{2017HEAD...2771250}. First results could come out in the next several months \citep{2017HEAD...1610404B}.

The NICER team will also study bursting sources, including 4U1636-536 \citep{2017HEAD...1610405M}. However, any $(M,R)$ information from this type of source will likely take considerably longer to generate from the data. That is because, unlike the expectations for non-accreting sources, analyses of bursting sources will require many separate data sets (a few per burst). 

Further in the future, third generation gravitational wave detectors such as the Einstein Telescope \citep{2010CQGra..27s4002P} or the Cosmic Explorer \citep{2017CQGra..34d4001A} could probe tidal effects in the gravitational wave signal of the inspiral of binary neutron star mergers, also providing information on the equation of state \citep{Hinderer:2009ca}. (If the binary is highly eccentric, even characteristic modes of oscillation could be excited during the inspiral with enough amplitude for detection \citep{2017ApJ...837...67C}, and universal relations could then be used to solve the inverse problem \citep{Chirenti:2015dda}.) 

Preliminary constraints on the tidal deformability parameters of the neutron star have already been recently proposed by the analysis of the first binary neutron star merger event, GW170817, and seem to favor softer equations of state (see fig. 5 of \citealt{PhysRevLett.119.161101}).

\section{Accurate models for the r-mode frequencies}
\label{sec:rmode}

\subsection{Estimates for $M$ and $R$ from the r-mode frequencies}

Here we will follow a different route from Section \ref{sec:M-R} and try to obtain constraints on the mass and radius of two neutron stars based on their (possibly) observed r-mode frequencies. If the frequencies reported by \citet{2014ApJ...784...72S,2014ApJ...793L..38S} indeed corresponded to an r-mode, we can improve the approach used by \citet{2014MNRAS.442.1786A} to estimate their masses and radii. In our previous work \citep{2017PhRvD..95f4060J}, we solved the linear perturbation equations in the Cowling approximation and obtained for the $\ell = m = 2$ r-mode frequency in the rotating frame, $\sigma_R$,
\begin{equation}
  \frac{\sigma_R}{\Omega}\equiv\kappa = \kappa_0 + \kappa_2\, \frac{\Omega^2}{\pi G \bar{\rho}} + \kappa_3\,,
  \label{eq:kappaxdiffrot}
\end{equation}
where $\Omega = 2\pi f$ is the angular velocity of the star, $\bar{\rho}$ is the average density of the star, $\kappa_0$ is the value in the slow rotation approximation, $\kappa_2$ is the rotational correction and $\kappa_3$ is a correction that arises if we allow the neutron star to have differential rotation, which we included in the form of the $j$-constant law \citep{1989MNRAS.237..355K,1989MNRAS.239..153K}. 

We found in \citep{2017PhRvD..95f4060J} that the $\kappa_i$ corrections to $\kappa$ are functions of the compactness $M/R$, and we expect them to be approximately independent of the equation of state \citep{2015PhRvD..91b4001I}.

Therefore, if $\sigma_R$ and $\Omega$ are both determined observationally, we can use a representative range for the neutron star radius to solve eq. (\ref{eq:kappaxdiffrot}), or an improved version of it, for the mass of the star.

\subsection{Going beyond the Cowling approximation}
\label{sec:Cowling}

The most obvious and important improvement that is needed is to go beyond the Cowling approximation. In \citet{2017PhRvD..95f4060J} we presented a first direct computation of the error in the r-mode frequencies due to the Cowling approximation. This was done by extrapolating our numerical results, obtained for fast rotating relativistic polytropic stars in the Cowling approximation, to $f \to 0$ (which gives the $\kappa_0$ term in eq.(\ref{eq:kappaxdiffrot})) and comparing with the fully relativistic (non-Cowling) results calculated in the slow-rotation approximation (see eq.(70) from \citealt{2015PhRvD..91b4001I}). We found that the error increases with compactness, ranging from 6\% to 11\% in the models we considered. 

\begin{figure}
\includegraphics[width=0.8\columnwidth]{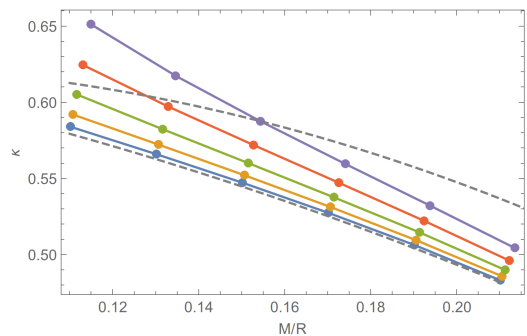}
\caption{$\kappa(M/R)$ curves for $f=$ 100, 200, 300, 400 and 500 Hz, from bottom to top (solid curves). The lower  dashed curve is the extrapolation of our results for $f \to 0$ and from it we find $\kappa_0 = 0.644 - 0.385 (M/R) - 1.84 (M/R)^2$. 
The upper dashed curve shows the fully relativistic $\kappa_0^{\textrm{GR}}$ given in eq.(70) from \citet{2015PhRvD..91b4001I} 
and we take the difference between the two dashed curves as our $\kappa_{\textrm{corr}}$ defined in 
eq.(\ref{eq:corr}). 
}
\label{fig1}
\end{figure}

In Fig. \ref{fig1} we present a first improvement in our extrapolation for the slow-rotation case, given by the lower dashed curve, where we used mode recycling to get more precise results for the r-mode frequencies. The upper dashed curve shows $\kappa_0$ calculated in full general relativity by \citet{2015PhRvD..91b4001I}. The difference between the two dashed curves gives the error due to the Cowling approximation in the slow-rotation limit.

So far there is no code available that can do fast rotation and fully relativistic simulations because of the coupling in the perturbation equations. Our error estimate is reliable for slowly rotating models, but what about neutron stars with faster  rotation? In order to answer this question, we turn to \citet{1997MNRAS.289..117Y}, where Yoshida and Kojima present a study on the accuracy of the Cowling approximation for f-modes. They write the f-mode frequency $\sigma$ of a rotating star as
\begin{equation}
\sigma = \sigma_0 + \sigma_0'm\epsilon\,,
\end{equation}
where $\sigma_0$ ($\sigma_R$ in the notation of eq. (21) in \citealt{1997MNRAS.289..117Y}) is the real part of the frequency of the f-mode for the non-rotating star, and the second term is a parametrization of the rotational correction to the $l = m$ f-mode frequency, in terms of the azimuthal number $m$ and $\epsilon = \Omega/\sqrt{M/R^3}$ (angular velocity of the star normalized by the Keplerian break-up speed).

A comparison between the $\sigma_0$ and $\sigma_0'$ obtained with and without the Cowling approximation is presented in figs. 1a and 1b of \citet{1997MNRAS.289..117Y}. They find that (a) the error due to the Cowling approximation only depends on $M/R$; (b) the behavior of the rotational correction $\sigma_0'$ is the same for Cowling and non-Cowling calculations; and (c) the error due to the Cowling approximation in $\sigma_0'$ is small. For f-modes, they find that the maximum relative error in $\sigma_0'$  is $\approx 1.9\%$ (for $l = 2$ and $M/R = 0.05$). For p-modes, the maximum relative error is $\approx 1.1\%$.

We must stress that the rotational corrections were calculated in the slow-rotation approximation in  \citet{1997MNRAS.289..117Y}. However, our analysis of their results gives us confidence to use our error estimate for the frequencies (shown in Fig. \ref{fig1}) also for our faster rotating stars. Therefore we propose to formally substitute the first term in eq.(\ref{eq:kappaxdiffrot}) by
\begin{equation}
\kappa_0 \to \kappa_0 + \kappa_{\textrm{corr}} \equiv \kappa_0^{\textrm{GR}}\,,
\label{eq:corr}
\end{equation}
with $\kappa_0$, $\kappa_{\textrm{corr}}$ and $\kappa_0^{\textrm{GR}}$ given as in the caption of Fig. \ref{fig1}.
As the Cowling approximation is more accurate for r-modes than it is for f-modes, we conservatively estimate that the error we are making for our fastest rotating stars should be lower that 5\%.

\subsection{Rotational correction}
\label{sec:rot}

The expression we obtained for the rotational $\kappa_2$ as a function of $M/R$ in eq.(\ref{eq:kappaxdiffrot}) in \citet{2017PhRvD..95f4060J} was calculated for different polytropes with increasing compactness and fixed volumetric radius $R = 14.15$ km from a fit $\kappa = a + bf^2$ of our numerical data. In order to use our results for different values of $R$, we performed now a few simulations with fixed $M/R$ and we found that $b \propto R^2$. This resembles the functional dependence of the second term in eq.(\ref{eq:kappaxdiffrot}), which can be explicitly written as
\begin{equation}
\kappa_2\frac{4\Omega^2}{3G}\frac{R^2}{M/R}\,,
\end{equation}
and it also agrees with eq. (4.11) of \cite{1999PhRvD..60f4006L} for Newtonian stars (including the perturbation of the Newtonian potential)
\footnote{Note that in \citet{1999PhRvD..60f4006L} $R$ denotes the radius of the corresponding non-rotating model. We found that the volumetric radius can be used to generalize their expressions in the cases of fastly spinning and noticeably deformed stars.}. 

\begin{figure}[!htb]
\begin{center}
\includegraphics[width=0.80\columnwidth]{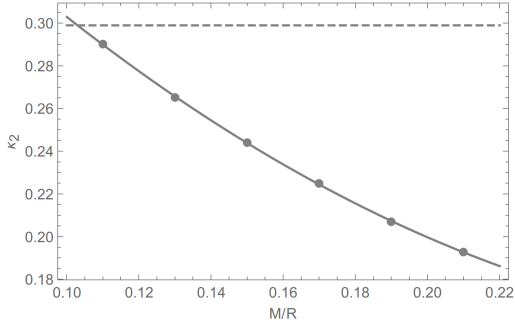}
\end{center}
\caption{The rotational correction $\kappa_2$ as a function of the compactness $M/R$. The solid line shows our results obtained in the relativistic Cowling approximation, from which we find $\kappa_2 = 0.466 - 1.93  (MR) + 2.98 (MR)^2$. The dashed line shows the corresponding value $\kappa_2 = 0.299$ obtained for the same stars, but this time using eq. (4.11) of \citet{1999PhRvD..60f4006L}, which uses the two potential formalism that includes the perturbation of the Newtonian potential.}
\label{fig2}
\end{figure}

We have compared our results for $\kappa_2$ with the formula given by \citet{1999PhRvD..60f4006L}, see Fig. \ref{fig2}. Our results are nearly coincident for lower compactness (and reproduce the value for the $n = 1$ Newtonian polytropic star with $M/R = 0.165$ quoted in their Table I), but the difference increases toward higher compactness. This is consistent with the expectation for a comparison with the perturbation equations for Newtonian stars. It is also important to note that this is one of the key ingredients responsible for the difference between our mass estimate for XTE J1751-305 and the value predicted by \citet{2014MNRAS.442.1786A}, as can be seen in Fig. \ref{fig3} and discussed below.

\subsection{Differential rotation}

A small amount of differential rotation could be present in neutron stars in LMXBs, and it might be relevant for the determination of the r-mode frequencies in stars with high compactness \citep{2017PhRvD..95f4060J}. More specifically, r-modes (as well as accretion) could be responsible for generating differential rotation in a neutron star, but viscosity in general and magnetic braking in particular could quickly bring the star back to uniform rotation \citep{2001PhRvD..64j4013R,2001PhRvD..64j4014R}.  So how much differential rotation should be expected?  It is believed that differential rotation can play an important role in the glitch phenomenon, and in sec. 4.3 of \citet{2014ApJ...789..141L} it is suggested that the maximum velocity difference that could be sustained in the inner crust due to vortex pinning could be $\approx r\omega_{\textrm{crit}} \approx 10^6 \textrm{cm s}^{-1}$, and it would produce a variation $\Delta f \approx 1$ Hz inside the star. 

The usual parametrization of the $j$-const. law for differential rotation is given in terms of the dimensionless quantity $\hat{A}$ \citep{2000ApJ...528L..29B}. However, differentially rotating stars with the same $\hat{A}$ but different compactness will have different values for $\Delta f$ \citep{2017PhRvD..95f4060J}, which could be considered as a more realistic measure of the differential rotation of the star. 

In \cite{2017PhRvD..95f4060J} we found a parametrization of the differential rotation correction $\kappa_3$ for neutron stars with strong differential rotation and $\hat{A}$ in the range $[0,0.5]$. However, for a more realistic differential rotation with $\Delta f \lessapprox 10$ Hz, we must have $\hat{A} \lessapprox 0.1$ and in this range we find that $\kappa_3$ is best described by 
\begin{equation}
\kappa_3 = c\hat{A}^{-2} + d\hat{A}^{-4}
\end{equation}
where we find $c = 0.27$ and $d = 0.33$ on average for models with $f = 400,\, 500,\, 600$ Hz and $M/R = 0.17,\, 0.19,\, 0.21$ (this is the relevant range for XTE J1751-305 and 4U 1636-536, see Section \ref{sec:results} below for more details).

However, if we choose to describe $\kappa_3$ in terms  of $\Delta f$ instead of $\hat{A}$ in this range, we find to lowest order in $\Delta f/f$
\begin{equation} 
\kappa_3 \approx \left(-0.046 + 0.91\frac{M}{R}\right)\frac{\Delta f}{f}\,.
\end{equation}
This is the expression we will use in order to fix the physical differential rotation parameter $\Delta f$ in our results given in the next Section \ref{sec:results}.

\section{Numerical results for XTE J1751-305 and 4U 1636-536}
\label{sec:results}

Now we are in possession of the best estimate for the r-mode frequency as a function of the stellar compactness, radius and angular velocity, and we can use it to estimate the compactness of the two stars XTE J1751-305 and 4U 1636-536 for which it was reported that a frequency compatible with an r-mode was detected.

\citet{2014ApJ...784...72S} found an oscillation with a frequency of 0.5727597 times the spin frequency of 435 Hz in the 2002 discovery outburst of XTE J1751-305. An interpretation of this frequency as an r-mode therefore implies $\kappa = 0.5727597$. 

For the 4U 1636-536 star, \cite{2014ApJ...793L..38S} report that they detected a coherent modulation at a frequency of $835.6440\pm0.0002$ Hz, which is 1.43546 times the stellar spin frequency of 582.14323 Hz (in the inertial frequency frame). In this case, we find $\kappa = 2 - 835.6440 / 582.14323 = 0.5645388$.

\begin{figure}
\begin{center}
\includegraphics[width=0.80\columnwidth]{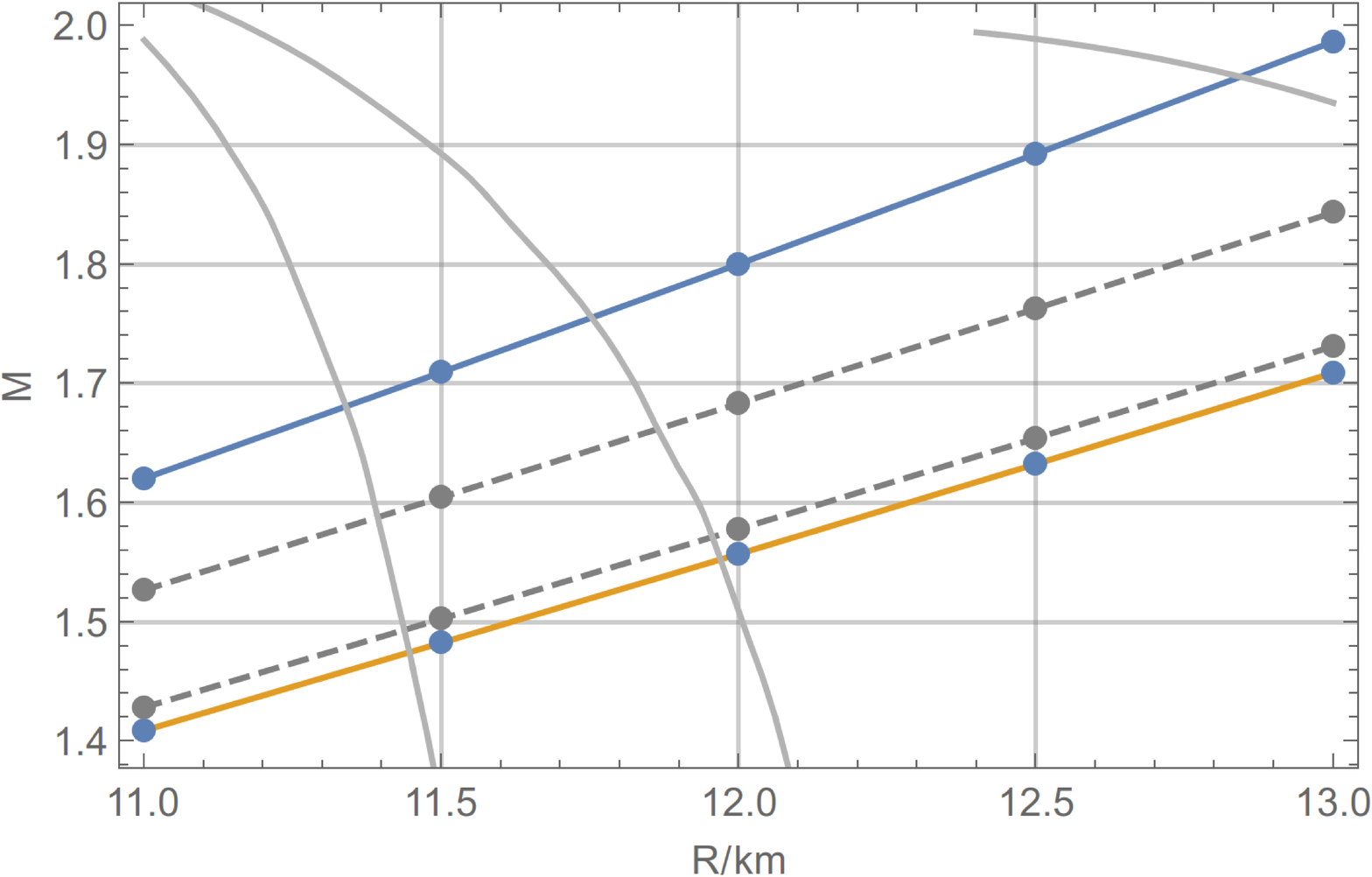}
\includegraphics[width=0.80\columnwidth]{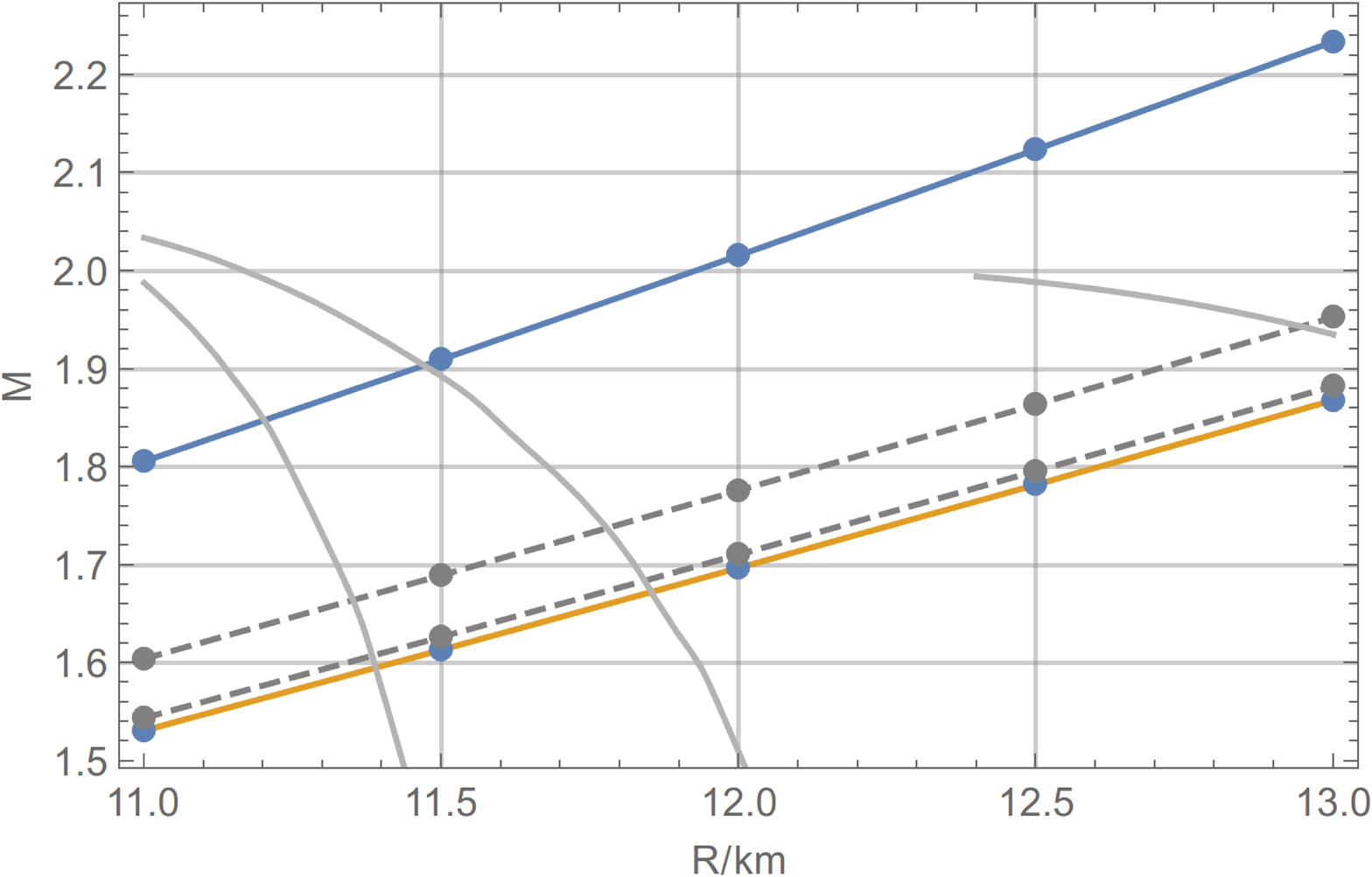}
\end{center}
\caption{Top: The blue line shows the constraint on $M/R$ for XTE J1751-305 from eq.(5) in \citet{2014MNRAS.442.1786A}, while the yellow line represents our result from eq.\ref{eq:kappaxdiffrot} for a uniformly rotating star ($\kappa_3 = 0$) with the improvements on $\kappa_0$, $\kappa_2$ and the correction due to the Cowling approximation discussed in Sections \ref{sec:Cowling} and \ref{sec:rot}. The dashed lines show the effect of a stronger differential rotation with $\Delta f = 10,\, 50$ Hz for better visualization and the transversal curves show $M(R)$ for three realistic tabulated equations of state (from left to right: APR4 \citealt{1998PhRvC..58.1804A}, SLy \citealt{2001A&A...380..151D} and H4 \citealt{2006PhRvD..73b4021L}).
Bottom: Same as the top plot, but for 4U 1636-536.}
\label{fig3}
\end{figure}

With these values of $\kappa$ for both stars and a representative range for the neutron star radius (see Section \ref{sec:M-Rdata}), we can solve for $M$ in eq.(\ref{eq:kappaxdiffrot}), including the $\kappa_{\textrm{corr}}$ term from eq.(\ref{eq:corr}), the rotational correction $\kappa_2$ as given by Fig. \ref{fig2} and a differential rotation term $\kappa_3$ consistent with a realistic differential rotation with $\Delta f \lessapprox 1$ Hz inside the star. In Fig. \ref{fig3} we present our results for uniformly and differentially rotating models, and compare with the fit given by eq.(5) of \citet{2014MNRAS.442.1786A}. 

Our results for uniformly rotating stars are $\approx 15\%$ lower than the estimate given by \citet{2014MNRAS.442.1786A}, where the fully relativistic r-mode frequency for slowly rotating uniform density stars from \citet{2003PhRvD..68l4010L} was combined with the second order rotational corrections for Newtonian polytropes from \citet{1999PhRvD..60f4006L}. As we mentioned in Section \ref{sec:rot}, the main factor for this difference is our improved calculation of the rotational correction $\kappa_2$ (see Fig. \ref{fig2}).
Differential rotation increases our mass estimate, but the effect is rather small: $\Delta f = 10$ Hz increases the total mass by only $\approx 0.02 M_{\odot}$.

In the range $11-13$ km for the stellar radius, we find from our results $1.41-1.71 M_{\odot}$ for XTE J1751-305 and $1.53-1.87 M_{\odot}$ for 4U 1636-536.
We can use the transversal curves in Fig. \ref{fig3}, which indicate $M(R)$ for a choice of three realistic tabulated equations of state (APR4  \citealt{1998PhRvC..58.1804A}, SLy \citealt{2001A&A...380..151D} and H4 \citealt{2006PhRvD..73b4021L}) to gain more insight. Each equation of state crosses our $M(r)$ line only once, providing a unique prediction for $M$ and $R$. However, we can see that $H4$ requires $R > 13$ km for both stars and is therefore less favored according to current expectations for the neutron star radius. 

Combining the constraints from both the r-mode frequencies and the APR4 and SLy equations of state, we find a more restricted range of values for $M$ and $R$: $R \approx 11.5-12$ km and $M \approx 1.48 - 1.56 M_{\odot}$ for XTE J1751-305 and $R \approx 11.4-11.9$ km and $M \approx 1.60 - 1.68 M_{\odot}$ for 4U 1636-536. However, it must be stressed here that these tight constraints are only valid if the frequencies reported in \citep{2014ApJ...784...72S,2014ApJ...793L..38S} are indeed core r-mode frequencies (see the discussion in Section \ref{sec:intro}).

\section{Conclusions}
\label{sec:conclusion}
 
Here we have explored the consequences for the mass and radius estimates of these two stars if the frequencies detected by Mahmoodifar and Strohmayer really correspond to their r-modes. We used a careful calculation of the r-mode frequency, taking into account relativistic, rotational and differential rotation corrections, as well as an estimate for the error due to the Cowling approximation, to find improved constraints on the compactness $M/R$ of XTE J1751-305 and 4U 1636-536. Our results yield, for a representative radius of 12 km, masses of $1.55 M_{\odot}$ and $1.70 M_{\odot}$, respectively. For XTE J1751-305, our value is approximately 15\% lower than what was obtained by \citet{2014MNRAS.442.1786A}, and we found no other estimate for the mass of 4U 1636-536 in the literature.

Differential rotation has a small effect on the mass estimates. If we allow the neutron star to have a $\Delta f$ of 10 Hz, this produces an offset in the mass of approximately $0.02 M_{\odot}$. This effect seems to be roughly linear in $\Delta f$, so a more realistic differential rotation of 1 Hz would be virtually undetectable.

We also analyzed three different realistic tabulated equations of state combined with our constraints on $M/R$ from the r-mode frequencies. Our analysis favors more compact models. For both of our two stars we found that H4 requires  $R > 13$ km, which is out of our representative range of $\approx$ 11 to $\approx$ 13 km \citep{2016EPJA...52...63M}. 

We used the APR4 and SLy equations of state to refine our predictions for mass and radius at the end of Section \ref{sec:results}, and our final mass estimates are $M \approx 1.48 - 1.56 M_{\odot}$ for XTE J1751-305 and $M \approx 1.60 - 1.68 M_{\odot}$ for 4U 1636-536.

Our results can be used as an indication in favor of softer equations of state like APR4 and SLy, but they should be taken with care, especially because of the disputed nature of the (possible) r-mode frequencies that we have analyzed here. Ideally, other observations including mass and radius estimates for these stars should be compared with our results to provide more conclusive evidence, also to prove or disprove the r-mode picture as an explanation for the observed frequencies.

\section*{Acknowledgements}

We would like to thank Ian Jones, Simin Mahmoodifar and Cole Miller for useful comments and also Lee Lindblom and Gregory Mendell for their help with the implementation of their equation for $\kappa_2$. This work was supported by the S\~ao Paulo Research Foundation (FAPESP grant 2015/20433-4).




\bibliographystyle{mnras}
\bibliography{draft} 

\begin{thebibliography}{}
\makeatletter
\relax
\def\mn@urlcharsother{\let\do\@makeother \do\$\do\&\do\#\do\^\do\_\do\%\do\~}
\def\mn@doi{\begingroup\mn@urlcharsother \@ifnextchar [ {\mn@doi@}
  {\mn@doi@[]}}
\def\mn@doi@[#1]#2{\def\@tempa{#1}\ifx\@tempa\@empty \href
  {http://dx.doi.org/#2} {doi:#2}\else \href {http://dx.doi.org/#2} {#1}\fi
  \endgroup}
\def\mn@eprint#1#2{\mn@eprint@#1:#2::\@nil}
\def\mn@eprint@arXiv#1{\href {http://arxiv.org/abs/#1} {{\tt arXiv:#1}}}
\def\mn@eprint@dblp#1{\href {http://dblp.uni-trier.de/rec/bibtex/#1.xml}
  {dblp:#1}}
\def\mn@eprint@#1:#2:#3:#4\@nil{\def\@tempa {#1}\def\@tempb {#2}\def\@tempc
  {#3}\ifx \@tempc \@empty \let \@tempc \@tempb \let \@tempb \@tempa \fi \ifx
  \@tempb \@empty \def\@tempb {arXiv}\fi \@ifundefined
  {mn@eprint@\@tempb}{\@tempb:\@tempc}{\expandafter \expandafter \csname
  mn@eprint@\@tempb\endcsname \expandafter{\@tempc}}}

\bibitem[\protect\citeauthoryear{Abbott et~al.}{Abbott
  et~al.}{2016a}]{PhysRevLett.116.061102}
Abbott B.~P.,  et~al., 2016a, \mn@doi [Phys. Rev. Lett.]
  {10.1103/PhysRevLett.116.061102}, 116, 061102

\bibitem[\protect\citeauthoryear{Abbott et~al.}{Abbott
  et~al.}{2016b}]{PhysRevLett.116.241103}
Abbott B.~P.,  et~al., 2016b, \mn@doi [Phys. Rev. Lett.]
  {10.1103/PhysRevLett.116.241103}, 116, 241103

\bibitem[\protect\citeauthoryear{{Abbott} et~al.,}{{Abbott}
  et~al.}{2017a}]{2017CQGra..34d4001A}
{Abbott} B.~P.,  et~al., 2017a, \mn@doi [Classical and Quantum Gravity]
  {10.1088/1361-6382/aa51f4}, \href
  {http://adsabs.harvard.edu/abs/2017CQGra..34d4001A} {34, 044001}

\bibitem[\protect\citeauthoryear{Abbott et~al.}{Abbott
  et~al.}{2017b}]{PhysRevLett.118.221101}
Abbott B.~P.,  et~al., 2017b, \mn@doi [Phys. Rev. Lett.]
  {10.1103/PhysRevLett.118.221101}, 118, 221101

\bibitem[\protect\citeauthoryear{Abbott et~al.}{Abbott
  et~al.}{2017c}]{PhysRevLett.119.141101}
Abbott B.~P.,  et~al., 2017c, \mn@doi [Phys. Rev. Lett.]
  {10.1103/PhysRevLett.119.141101}, 119, 141101

\bibitem[\protect\citeauthoryear{Abbott et~al.}{Abbott
  et~al.}{2017d}]{PhysRevLett.119.161101}
Abbott B.~P.,  et~al., 2017d, \mn@doi [Phys. Rev. Lett.]
  {10.1103/PhysRevLett.119.161101}, 119, 161101

\bibitem[\protect\citeauthoryear{{Akmal}, {Pandharipande}  \&
  {Ravenhall}}{{Akmal} et~al.}{1998}]{1998PhRvC..58.1804A}
{Akmal} A.,  {Pandharipande} V.~R.,   {Ravenhall} D.~G.,  1998, \mn@doi [\prc]
  {10.1103/PhysRevC.58.1804}, \href
  {http://adsabs.harvard.edu/abs/1998PhRvC..58.1804A} {58, 1804}

\bibitem[\protect\citeauthoryear{{Andersson}, {Jones}  \& {Ho}}{{Andersson}
  et~al.}{2014}]{2014MNRAS.442.1786A}
{Andersson} N.,  {Jones} D.~I.,   {Ho} W.~C.~G.,  2014, \mn@doi [\mnras]
  {10.1093/mnras/stu870}, \href
  {http://adsabs.harvard.edu/abs/2014MNRAS.442.1786A} {442, 1786}

\bibitem[\protect\citeauthoryear{{Arzoumanian}, {Gendreau}  \&
  {Markwardt}}{{Arzoumanian} et~al.}{2017}]{2017HEAD...2771250}
{Arzoumanian} Z.,  {Gendreau} K.,   {Markwardt} C.,  2017, in AAS/High Energy
  Astrophysics Division.

\bibitem[\protect\citeauthoryear{{Baumgarte}, {Shapiro}  \&
  {Shibata}}{{Baumgarte} et~al.}{2000}]{2000ApJ...528L..29B}
{Baumgarte} T.~W.,  {Shapiro} S.~L.,   {Shibata} M.,  2000, \mn@doi [\apjl]
  {10.1086/312425}, \href {http://adsabs.harvard.edu/abs/2000ApJ...528L..29B}
  {528, L29}

\bibitem[\protect\citeauthoryear{{Bogdanov} et~al.,}{{Bogdanov}
  et~al.}{2017}]{2017HEAD...1610404B}
{Bogdanov} S.,  et~al., 2017, in AAS/High Energy Astrophysics Division. p.
  104.04

\bibitem[\protect\citeauthoryear{{Chandrasekhar}}{{Chandrasekhar}}{1970}]{1970ApJ...161..561C}
{Chandrasekhar} S.,  1970, \mn@doi [\apj] {10.1086/150560}, \href
  {http://adsabs.harvard.edu/abs/1970ApJ...161..561C} {161, 561}

\bibitem[\protect\citeauthoryear{{Chirenti} \& {Sk{\'a}kala}}{{Chirenti} \&
  {Sk{\'a}kala}}{2013}]{2013PhRvD..88j4018C}
{Chirenti} C.,  {Sk{\'a}kala} J.,  2013, \mn@doi [\prd]
  {10.1103/PhysRevD.88.104018}, \href
  {http://adsabs.harvard.edu/abs/2013PhRvD..88j4018C} {88, 104018}

\bibitem[\protect\citeauthoryear{{Chirenti}, {Sk{\'a}kala}  \&
  {Yoshida}}{{Chirenti} et~al.}{2013}]{2013PhRvD..87d4043C}
{Chirenti} C.,  {Sk{\'a}kala} J.,   {Yoshida} S.,  2013, \mn@doi [\prd]
  {10.1103/PhysRevD.87.044043}, \href
  {http://adsabs.harvard.edu/abs/2013PhRvD..87d4043C} {87, 044043}

\bibitem[\protect\citeauthoryear{Chirenti, de Souza  \& Kastaun}{Chirenti
  et~al.}{2015}]{Chirenti:2015dda}
Chirenti C.,  de Souza G.~H.,   Kastaun W.,  2015, \mn@doi [Phys. Rev.]
  {10.1103/PhysRevD.91.044034}, D91, 044034

\bibitem[\protect\citeauthoryear{{Chirenti}, {Gold}  \& {Miller}}{{Chirenti}
  et~al.}{2017}]{2017ApJ...837...67C}
{Chirenti} C.,  {Gold} R.,   {Miller} M.~C.,  2017, \mn@doi [\apj]
  {10.3847/1538-4357/aa5ebb}, \href
  {http://adsabs.harvard.edu/abs/2017ApJ...837...67C} {837, 67}

\bibitem[\protect\citeauthoryear{{Douchin} \& {Haensel}}{{Douchin} \&
  {Haensel}}{2001}]{2001A&A...380..151D}
{Douchin} F.,  {Haensel} P.,  2001, \mn@doi [\aap]
  {10.1051/0004-6361:20011402}, \href
  {http://adsabs.harvard.edu/abs/2001A%26A...380..151D} {380, 151}

\bibitem[\protect\citeauthoryear{{Friedman} \& {Schutz}}{{Friedman} \&
  {Schutz}}{1978}]{1978ApJ...222..281F}
{Friedman} J.~L.,  {Schutz} B.~F.,  1978, \mn@doi [\apj] {10.1086/156143},
  \href {http://adsabs.harvard.edu/abs/1978ApJ...222..281F} {222, 281}

\bibitem[\protect\citeauthoryear{{Gandolfi}, {Carlson}  \& {Reddy}}{{Gandolfi}
  et~al.}{2012}]{2012PhRvC..85c2801G}
{Gandolfi} S.,  {Carlson} J.,   {Reddy} S.,  2012, \mn@doi [\prc]
  {10.1103/PhysRevC.85.032801}, \href
  {http://adsabs.harvard.edu/abs/2012PhRvC..85c2801G} {85, 032801}

\bibitem[\protect\citeauthoryear{Hinderer, Lackey, Lang  \& Read}{Hinderer
  et~al.}{2010}]{Hinderer:2009ca}
Hinderer T.,  Lackey B.~D.,  Lang R.~N.,   Read J.~S.,  2010, \mn@doi [Phys.
  Rev.] {10.1103/PhysRevD.81.123016}, D81, 123016

\bibitem[\protect\citeauthoryear{{Idrisy}, {Owen}  \& {Jones}}{{Idrisy}
  et~al.}{2015}]{2015PhRvD..91b4001I}
{Idrisy} A.,  {Owen} B.~J.,   {Jones} D.~I.,  2015, \mn@doi [\prd]
  {10.1103/PhysRevD.91.024001}, \href
  {http://adsabs.harvard.edu/abs/2015PhRvD..91b4001I} {91, 024001}

\bibitem[\protect\citeauthoryear{{Jasiulek} \& {Chirenti}}{{Jasiulek} \&
  {Chirenti}}{2017}]{2017PhRvD..95f4060J}
{Jasiulek} M.,  {Chirenti} C.,  2017, \mn@doi [\prd]
  {10.1103/PhysRevD.95.064060}, \href
  {http://adsabs.harvard.edu/abs/2017PhRvD..95f4060J} {95, 064060}

\bibitem[\protect\citeauthoryear{{Komatsu}, {Eriguchi}  \& {Hachisu}}{{Komatsu}
  et~al.}{1989a}]{1989MNRAS.237..355K}
{Komatsu} H.,  {Eriguchi} Y.,   {Hachisu} I.,  1989a, \mn@doi [\mnras]
  {10.1093/mnras/237.2.355}, \href
  {http://adsabs.harvard.edu/abs/1989MNRAS.237..355K} {237, 355}

\bibitem[\protect\citeauthoryear{{Komatsu}, {Eriguchi}  \& {Hachisu}}{{Komatsu}
  et~al.}{1989b}]{1989MNRAS.239..153K}
{Komatsu} H.,  {Eriguchi} Y.,   {Hachisu} I.,  1989b, \mn@doi [\mnras]
  {10.1093/mnras/239.1.153}, \href
  {http://adsabs.harvard.edu/abs/1989MNRAS.239..153K} {239, 153}

\bibitem[\protect\citeauthoryear{{Lackey}, {Nayyar}  \& {Owen}}{{Lackey}
  et~al.}{2006}]{2006PhRvD..73b4021L}
{Lackey} B.~D.,  {Nayyar} M.,   {Owen} B.~J.,  2006, \mn@doi [\prd]
  {10.1103/PhysRevD.73.024021}, \href
  {http://adsabs.harvard.edu/abs/2006PhRvD..73b4021L} {73, 024021}

\bibitem[\protect\citeauthoryear{{Lattimer} \& {Steiner}}{{Lattimer} \&
  {Steiner}}{2014a}]{2014EPJA...50...40L}
{Lattimer} J.~M.,  {Steiner} A.~W.,  2014a, \mn@doi [European Physical Journal
  A] {10.1140/epja/i2014-14040-y}, \href
  {http://adsabs.harvard.edu/abs/2014EPJA...50...40L} {50, 40}

\bibitem[\protect\citeauthoryear{{Lattimer} \& {Steiner}}{{Lattimer} \&
  {Steiner}}{2014b}]{2014ApJ...784..123L}
{Lattimer} J.~M.,  {Steiner} A.~W.,  2014b, \mn@doi [\apj]
  {10.1088/0004-637X/784/2/123}, \href
  {http://adsabs.harvard.edu/abs/2014ApJ...784..123L} {784, 123}

\bibitem[\protect\citeauthoryear{{Lee}}{{Lee}}{2014}]{2014MNRAS.442.3037L}
{Lee} U.,  2014, \mn@doi [\mnras] {10.1093/mnras/stu1077}, \href
  {http://adsabs.harvard.edu/abs/2014MNRAS.442.3037L} {442, 3037}

\bibitem[\protect\citeauthoryear{{Lindblom}, {Mendell}  \& {Owen}}{{Lindblom}
  et~al.}{1999}]{1999PhRvD..60f4006L}
{Lindblom} L.,  {Mendell} G.,   {Owen} B.~J.,  1999, \mn@doi [\prd]
  {10.1103/PhysRevD.60.064006}, \href
  {http://adsabs.harvard.edu/abs/1999PhRvD..60f4006L} {60, 064006}

\bibitem[\protect\citeauthoryear{{Link}}{{Link}}{2014}]{2014ApJ...789..141L}
{Link} B.,  2014, \mn@doi [\apj] {10.1088/0004-637X/789/2/141}, \href
  {http://adsabs.harvard.edu/abs/2014ApJ...789..141L} {789, 141}

\bibitem[\protect\citeauthoryear{{Lockitch}, {Friedman}  \&
  {Andersson}}{{Lockitch} et~al.}{2003}]{2003PhRvD..68l4010L}
{Lockitch} K.~H.,  {Friedman} J.~L.,   {Andersson} N.,  2003, \mn@doi [\prd]
  {10.1103/PhysRevD.68.124010}, \href
  {http://adsabs.harvard.edu/abs/2003PhRvD..68l4010L} {68, 124010}

\bibitem[\protect\citeauthoryear{{Mahmoodifar} \& {Strohmayer}}{{Mahmoodifar}
  \& {Strohmayer}}{2017a}]{2017HEAD...1610405M}
{Mahmoodifar} S.,  {Strohmayer} T.~E.,  2017a, in AAS/High Energy Astrophysics
  Division. p. 104.05

\bibitem[\protect\citeauthoryear{{Mahmoodifar} \& {Strohmayer}}{{Mahmoodifar}
  \& {Strohmayer}}{2017b}]{2017ApJ...840...94M}
{Mahmoodifar} S.,  {Strohmayer} T.,  2017b, \mn@doi [\apj]
  {10.3847/1538-4357/aa6d62}, \href
  {http://adsabs.harvard.edu/abs/2017ApJ...840...94M} {840, 94}

\bibitem[\protect\citeauthoryear{{Miller} \& {Lamb}}{{Miller} \&
  {Lamb}}{2016}]{2016EPJA...52...63M}
{Miller} M.~C.,  {Lamb} F.~K.,  2016, \mn@doi [European Physical Journal A]
  {10.1140/epja/i2016-16063-8}, \href
  {http://adsabs.harvard.edu/abs/2016EPJA...52...63M} {52, 63}

\bibitem[\protect\citeauthoryear{{{\"O}zel} \& {Freire}}{{{\"O}zel} \&
  {Freire}}{2016}]{2016ARA&A..54..401O}
{{\"O}zel} F.,  {Freire} P.,  2016, \mn@doi [\araa]
  {10.1146/annurev-astro-081915-023322}, \href
  {http://adsabs.harvard.edu/abs/2016ARA%26A..54..401O} {54, 401}

\bibitem[\protect\citeauthoryear{{Punturo} et~al.}{{Punturo}
  et~al.}{2010}]{2010CQGra..27s4002P}
{Punturo} M.,  et~al., 2010, \mn@doi [Classical and Quantum Gravity]
  {10.1088/0264-9381/27/19/194002}, \href
  {http://adsabs.harvard.edu/abs/2010CQGra..27s4002P} {27, 194002}

\bibitem[\protect\citeauthoryear{{Rezzolla}, {Lamb}, {Markovi{\'c}}  \&
  {Shapiro}}{{Rezzolla} et~al.}{2001a}]{2001PhRvD..64j4013R}
{Rezzolla} L.,  {Lamb} F.~K.,  {Markovi{\'c}} D.,   {Shapiro} S.~L.,  2001a,
  \mn@doi [\prd] {10.1103/PhysRevD.64.104013}, \href
  {http://adsabs.harvard.edu/abs/2001PhRvD..64j4013R} {64, 104013}

\bibitem[\protect\citeauthoryear{{Rezzolla}, {Lamb}, {Markovi{\'c}}  \&
  {Shapiro}}{{Rezzolla} et~al.}{2001b}]{2001PhRvD..64j4014R}
{Rezzolla} L.,  {Lamb} F.~K.,  {Markovi{\'c}} D.,   {Shapiro} S.~L.,  2001b,
  \mn@doi [\prd] {10.1103/PhysRevD.64.104014}, \href
  {http://adsabs.harvard.edu/abs/2001PhRvD..64j4014R} {64, 104014}

\bibitem[\protect\citeauthoryear{{Strohmayer} \& {Mahmoodifar}}{{Strohmayer} \&
  {Mahmoodifar}}{2014a}]{2014ApJ...784...72S}
{Strohmayer} T.,  {Mahmoodifar} S.,  2014a, \mn@doi [\apj]
  {10.1088/0004-637X/784/1/72}, \href
  {http://adsabs.harvard.edu/abs/2014ApJ...784...72S} {784, 72}

\bibitem[\protect\citeauthoryear{{Strohmayer} \& {Mahmoodifar}}{{Strohmayer} \&
  {Mahmoodifar}}{2014b}]{2014ApJ...793L..38S}
{Strohmayer} T.,  {Mahmoodifar} S.,  2014b, \mn@doi [\apjl]
  {10.1088/2041-8205/793/2/L38}, \href
  {http://adsabs.harvard.edu/abs/2014ApJ...793L..38S} {793, L38}

\bibitem[\protect\citeauthoryear{{Yoshida} \& {Kojima}}{{Yoshida} \&
  {Kojima}}{1997}]{1997MNRAS.289..117Y}
{Yoshida} S.,  {Kojima} Y.,  1997, \mn@doi [\mnras] {10.1093/mnras/289.1.117},
  \href {http://adsabs.harvard.edu/abs/1997MNRAS.289..117Y} {289, 117}

\makeatother
\end{thebibliography}








\bsp	
\label{lastpage}
\end{document}